\documentclass[prl,raggedbottom,showpacs,twocolumn,superscriptaddress,floatfix]{revtex4}

\usepackage{graphicx}
\usepackage{amssymb}
\usepackage{amsmath}
\usepackage{amscd}
\usepackage{eucal}
\usepackage{color}
\usepackage{bm}

\newcommand{\e}{{\rm e}}

\newcommand{\half}{{\textstyle{\frac{1}{2}}}}

\newcommand{\eps}{\epsilon}

\newcommand{\eminus}{ e^{\operatorname{-}} }

\newcommand{\kB}{k_{\rm B}}

\newcommand{\ignore}[1]{\relax}

\definecolor{DarkGreen}{rgb}{0,0.7,0}

\begin{document}

\title{Most Efficient Quantum Thermoelectric at Finite Power Output}

\author{Robert S.~Whitney}
\affiliation{
Laboratoire de Physique et Mod\'elisation des Milieux Condens\'es (UMR 5493), 
Universit\'e Grenoble 1 and CNRS, Maison des Magist\`eres, BP 166, 38042 Grenoble, France.}

\date{February 21, 2014} 
\begin{abstract}

Machines are only Carnot efficient if they are reversible, but 
then their power output is {\it vanishingly small}.  
Here we ask, what is the maximum efficiency of an irreversible device with {\it finite} power output?
We use a nonlinear scattering theory to answer this question for thermoelectric quantum systems; 
heat engines or refrigerators consisting of nanostructures or molecules that exhibit a Peltier effect. 
We find that quantum mechanics places an upper bound on both power output, and on the efficiency at any finite power.  
The upper bound on efficiency equals Carnot efficiency at zero power output, 
but decays with increasing power output. It is intrinsically quantum (wavelength dependent),
unlike Carnot efficiency. 
This maximum efficiency occurs when the system lets through all particles in a certain energy window, but none at other energies.  
A physical implementation of this is discussed, as is the suppression of efficiency by a phonon heat flow.

\end{abstract}

\pacs{73.63.-b, 05.70.Ln,  72.15.Jf, 84.60.Rb}


\maketitle

{\bf Introduction.}
Quantum thermodynamics \cite{QuantumThermodyn-book} 
is the physics of thermodynamic processes in quantum systems, such as the conversion of heat to work.
This is of particular interest for the thermoelectric response
\cite{books,DiSalvo-review,Shakouri-reviews}
of nanostructures \cite{Pekola-reviews} or molecules 
\cite{Paulson-Datta2003,Reddy2007}.
It places fundamental bounds on the efficiency and power output of heat engines and refrigerators made 
from such systems,
such as Carnot's thermodynamic bound on efficiency
or Pendry's quantum bound 
on entropy flow \cite{Pendry1983}.

\begin{figure}[b]
\includegraphics[width=\columnwidth]{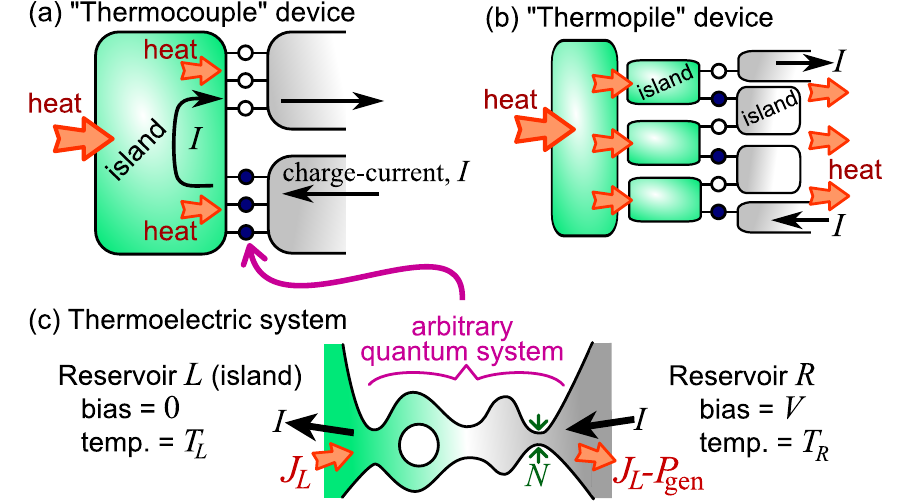}
\caption{\label{Fig:thermocouple} 
Typical thermoelectric devices are shown in (a) and (b),
with (c) showing the quantum system with the thermoelectric response.
To unify the analysis, 
heat is always taken to flow as shown, 
hence a heat engine has $T_L >T_R$, while a refrigerator has $T_L<T_R$.
In (a) and (b), the filled (open) circles are quantum systems where transport occurs via ``electron'' states above the Fermi surface (``hole'' states below the Fermi surface).
In (c), $N$ is the number of transverse modes in the narrowest part of the quantum system.
}
\end{figure}

The efficiencies of heat engines, $\eta_{\rm eng}$, and refrigerators, $\eta_{\rm fri}$,
are particularly important ($\eta_{\rm fri}$ is called the coefficient of performance, COP). 
These efficiencies are the ratio of power output to power input.
For a heat engine, the output is the electrical power, $P_{\rm gen}$,
and the input is the heat flow out of a reservoir (the left ($L$) reservoir in Fig.~\ref{Fig:thermocouple}c), $J_L$.
For a refrigerator, it is the inverse.  
For left ($L$) and right ($R$) reservoirs at temperatures $T_L$ and $T_R$,
Carnot's bounds on these efficiencies are
\begin{eqnarray}
\eta_{\rm eng}^{\rm Carnot} = 1-T_R/ T_L,  \quad \ 
\eta_{\rm fri}^{\rm Carnot} = (T_R/T_L -1)^{-1}, \ \  
\label{Eq:Carnot}
\end{eqnarray}
where heat flows as in Fig.~\ref{Fig:thermocouple},
so $T_L>T_R$ for heat engines and $T_R > T_L$ for refrigerators.
Proposals exist to achieve these efficiency  
in bulk \cite{Mahan-Sofo1996} or quantum 
\cite{Humphrey-Linke2005,Kim-Datta-Lundstrom2009,Jordan-Sothmann-Sanchez-Buttiker2013} systems. 

However Carnot efficiency is only achieved in 
{\it reversible systems}, which have vanishing power output.   
Any useful device must give a finite power output, and so be  
{\it irreversible}.
So what are the equivalents of Carnot efficiencies 
for such irreversible (entropy-producing) systems?
To be more precise, we note that engineers typically need a device to provide a certain power,
at the highest possible efficiency.
Thus we ask, what is the maximum allowed efficiency at  any {\it given} power output?
As physicists, we can also ask what is the least irreversible system 
(i.e.\ that which produces the least entropy) that delivers a 
 {\it given} power output?  With a little algebra, the first and second laws of thermodynamics \cite{Reichl-book} 
 tell us that 
a heat engine producing power $P$ must also produce entropy at a rate, 
 \begin{eqnarray}
\dot S (P) =  (P/T_R) \left({\eta^{\rm Carnot}_{\rm eng}/\eta_{\rm eng}} -1\right). 
\label{Eq:dotS-eng}
\end{eqnarray}
Similarly, a refrigerator with cooling power $J$ has
 \begin{eqnarray}
\dot S (J) =  (J/T_R) \left(1\big/\eta_{\rm fri} - 1\big/\eta^{\rm Carnot}_{\rm fri} \right). 
\label{Eq:dotS-fri}
\end{eqnarray}
Thus the two above questions are the same, since the most efficient system is the least irreversible.

{\bf Central results.}
We answer these questions for any thermoelectric quantum system 
that can be modeled with nonlinear Landauer-B\"uttiker scattering theory.

Firstly, we find that quantum mechanics places an upper bound on the power output of such systems,
\begin{eqnarray}
\hbox{Heat-engine:} & & \hskip -3mm P_{\rm gen} \,\leq\, P_{\rm gen}^{\rm QB2} \,\equiv\, 
 A_0\, {\pi^2 \over h} N \kB^2 \big(T_L-T_R\big)^2 \quad \quad
\label{Eq:P-qb2}
\\
\hbox{Refrigerator:} & & \hskip -3mm J_L \,\leq\, \half J_L^{\rm QB} \,\equiv\, {1 \over 12}\,{\pi^2 \over h} N \kB^2 T_L^2
\label{Eq:J-qb-fri}
\end{eqnarray}
where $A_0 \simeq 0.0321$. 
We refer to $P_{\rm gen}^{\rm QB2}$ and $J_L^{\rm QB}$ as quantum bounds (QB), as they depend on the number of transverse
modes in the quantum system, $N$, which scales like the inverse Fermi wavelength.
$J_L^{\rm QB}$ is 
Pendry's quantum bound on 
the heat current out of reservoir $L$  \cite{Pendry1983}. 
The ``2'' on $P_{\rm gen}^{\rm QB2}$ 
indicates that it is for {\it two}-lead systems \cite{footnote:2012w-2ndlaw}.

Secondly, we find a fundamental upper bound on the efficiencies at finite power output, which is {\it lower} than Carnot efficiency. 
The upper bound for a heat engine is a decaying function of 
$P_{\rm gen}/P_{\rm gen}^{\rm QB2}$, whereas the upper bound for a refrigerator is a decaying function of 
$J_L/J_L^{\rm QB}$. 
At small output power, these bounds on efficiencies are
\begin{eqnarray}
\eta_{\rm eng} \big(P_{\rm gen}\big) \,=\,  \eta_{\rm eng}^{\rm Carnot} 
\left(1- 0.478
\sqrt{  {T_R \over T_L} \ {P_{\rm gen} \over P_{\rm gen}^{\rm QB2}} }\ 
\right)\!, \quad
\label{Eq:eta-eng-small-Pgen}
\\
\eta_{\rm fri}(J_L) \,= \, \eta_{\rm fri}^{\rm Carnot} 
\left(1- 1.09
\sqrt{
\,{T_R \over T_R-T_L}\ {J_L \over J_L^{\rm QB}} }\, \right)\!,
\label{Eq:eta-fri-smallJ}
\end{eqnarray}
to lowest order in $P_{\rm gen}/P_{\rm gen}^{\rm QB2}$ and $J_L/J_L^{\rm QB}$, respectively.
In these limits, the least irreversible heat engine produces entropy at a rate $\dot S \propto P_{\rm gen}^{3/2}$,
while the least irreversible refrigerator does so at a rate 
$\dot S \propto J_L^{3/2}$.

\begin{figure}
\includegraphics[width=0.9\columnwidth]{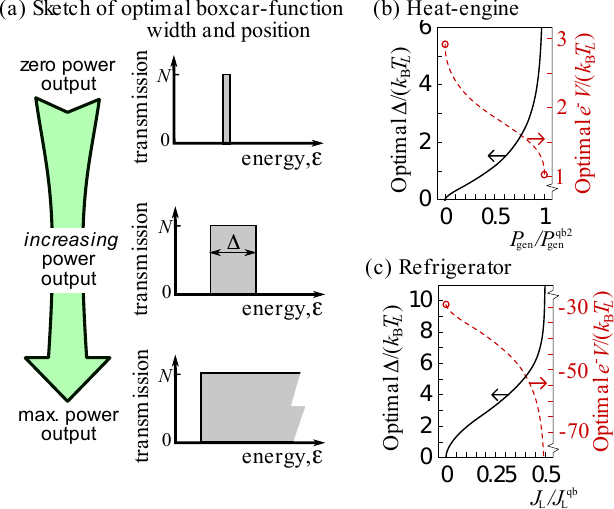}
\caption{\label{Fig:Delta+V}
(a) Sketch of how the optimal transmission changes as the required power output is increased.
Maximum power is when the right-hand edge of the boxcar function goes to $+\infty$.
The qualitative features always follow this sketch, while the quantitative details depend on
$T_R/T_L$.  
(b) Plots of optimal $\Delta$ (solid curve) and $\eminus V$ (dashed curve)
against heat engine power output, $P_{\rm gen}$,
at $T_R/T_L=0.1$. The first equation in 
Eq.~(\ref{Eq:eng-bounds}) then gives $\eps_0$.
(c) Plots of optimal $\Delta$ (solid) and $\eminus V$ (dashed) against cooling power, $J_L$, at $T_R/T_L=10$. The second equation in 
Eq.~(\ref{Eq:fri-bounds}) then gives $\eps_1$.
}
\end{figure}

These fundamental upper bounds on efficiencies at finite power
are of quantum origin (they are wavelength dependent), unlike Carnot's bounds 
(which were derived using classical physics).
They play the role for {\it irreversible} thermoelectric systems that Carnot's bounds 
do for {\it reversible} systems, and are more stringent than Carnot's bounds. 
This upper bound on efficiency is achieved when only particles in a given energy window (determined by the desired power output) {\color{red} traverse} the quantum system, see Fig.~\ref{Fig:Delta+V}a.   
Real systems will have lower efficiencies; 
improving them would only approach these bounds.

{\bf Nonlinear theory.}
Linear-response theory works in bulk systems for most $T_R/T_L$
 \cite{Footnote:bulk-semicond},  but a nonlinear theory is needed for quantum systems whenever 
$1-T_R/T_L$ is not small.
An example would be getting electricity from a thermoelectric between a diesel motor's exhaust 
$\simeq 700$K and its surroundings $\simeq 280$K (in which case 
the bound in Eq.~(\ref{Eq:P-qb2}) is $\sim$10 nW per transverse mode).

Interactions are crucial in the nonlinear regime,
and must be treated
in a manner appropriate to the system in question.
Here, we use a
nonlinear Landauer-B\"uttiker scattering formula, which
was first derived by treating electron-electron interactions as mean-field charging effects
\cite{Christen-ButtikerEPL96,Sanchez-Buttiker},
and recently applied to thermoelectric effects 
\cite{Sanchez-Lopez2013,2012w-pointcont,Meair-Jacquod2013,2012w-2ndlaw}.
Identical equations apply for resonant level models 
\cite{Galperin2007-2008,Humphrey-Linke2005,Murphy2008,Esposito2009-thermoelec,Nakpathomkun-Xu-Linke2010}, 
and have been derived from functional renormalization group \cite{Meden2013} for such models with 
single-electron charging effects.
Refs.~\cite{Bruneau2012,2012w-2ndlaw} show that such theories respect 
thermodynamics.
The heat current out of the $L$ reservoir into the quantum system, $J_L$, 
and the electrical power generated by the system, $P_{\rm gen} = V I_L$, 
are
\begin{eqnarray}
J_L  \! &=& \!
{1 \over h} \sum_\mu \int_0^\infty {\rm d}\eps 
\hskip 4mm \eps \hskip 5mm  
{\cal T}^{\mu\mu}_{RL}(\eps)   \ \big[f_L^\mu (\eps) - f_R^\mu (\eps)\big],
\label{Eq:JL}
\\
P_{\rm gen} \! &=& \!  
{1\over h} \sum_\mu \int_0^\infty {\rm d}\eps 
\ \, \mu\eminus V\ \,
{\cal T}^{\mu\mu}_{RL}(\eps)   \ \big[f_L^\mu (\eps) - f_R^\mu (\eps)\big], \ \ \  
\label{Eq:Pgen}
\end{eqnarray}
where $\eminus$ is the electron charge ($\eminus<0$).
The sum is over 
$\mu=1$ for ``electron'' states above the L reservoir's chemical potential,
and $\mu=-1$ for ``hole'' states below that chemical potential.
Interaction effects mean that the transmission function, ${\cal T}^{\mu\mu}_{RL}(\eps)$, 
 is a {\it self-consistently} determined function of $T_{L,R}$ and $V$.
The Fermi function for electrons entering
from reservoir $j$ is 
\begin{eqnarray}
f_j^\mu(\eps) = \left(1+\exp\left[(\eps - \mu \eminus V_j)\big/ (\kB T_j) \right] \right)^{-1}.
\nonumber
\end{eqnarray}

Scattering theory has been used to find the properties of 
many thermoelectric systems from their  ${\cal T}^{\mu\mu}_{RL}(\eps)$,
e.g.\ Refs.~\cite{Butcher1990,Molenkamp1992,Paulson-Datta2003,
Vavilov-Stone2005,
Zebarjadi2007,Galperin2007-2008,Murphy2008, Esposito2009-thermoelec, Nakpathomkun-Xu-Linke2010,
Nozaki2010,jw-epl,Saha2011,Karlstrom2011,jwmb,Jordan-Sothmann-Sanchez-Buttiker2013,
Sanchez-Lopez2013,2012w-pointcont,Meair-Jacquod2013,Meden2013,Hershfield2013}.
Here instead, we find the ${\cal T}^{\mu\mu}_{RL}(\eps)$ that maximizes efficiency
at given power output.
We initially assume only elastic scattering in the quantum system, 
although decoherence without relaxation is allowed as it does not change the structure of Eqs.~(\ref{Eq:JL},\ref{Eq:Pgen}).
Inelastic effects are briefly discussed at the end of this Letter.
We take each island (see Fig.~\ref{Fig:thermocouple}) 
as large enough to be a reservoir in local equilibrium.
This differs from the ``three-terminal'' systems
\cite{Entin-Wohlman2011,SB2011,SSJB2012,Trocha2012,Brandner2013},
in which particles remain coherent in the island.
Here, we only discuss
electrons dominating transmission 
(filled circles in Fig.~\ref{Fig:thermocouple}). 
When holes dominate
(open circles in Fig.~\ref{Fig:thermocouple}) one takes
${\cal T}_{RL}^{\mu,\mu} (\eps) \to {\cal T}_{RL}^{-\mu,-\mu} (\eps)$
with $V \to -V$, then $I_L \to -I_L$ while
$J_L$ and $P_{\rm gen}$ are unchanged.

\begin{figure}[b]
\includegraphics[width=\columnwidth]{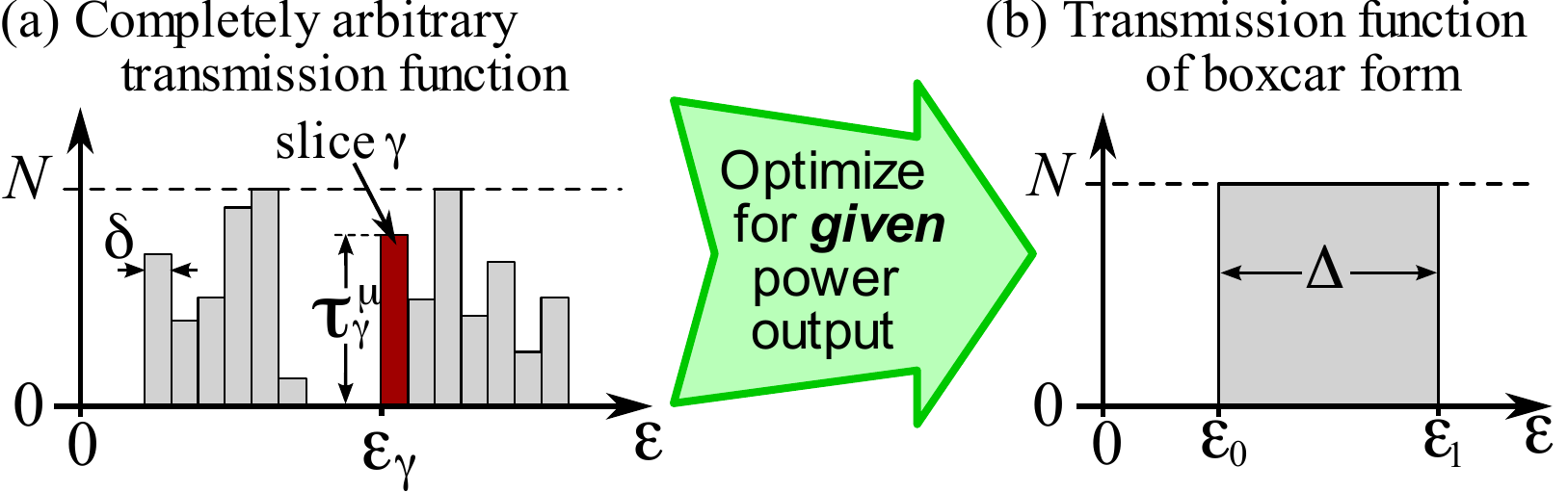}
\caption{\label{Fig:T-functions}
Finding the  $ {\cal T}_{RL}^{\mu\mu} (\eps)$ 
that maximizes the efficiency.
In (a) $ {\cal T}_{RL}^{\mu\mu} (\eps)$ is considered as 
infinitely many slices of width $\delta \to 0$, so slice $\gamma$ has energy 
$\eps_\gamma \equiv \gamma \delta$ and height $\tau^{\mu}_\gamma$. 
This gives (b) with a transcendental equation for $\eps_0$ and $\eps_1$.  }
\end{figure}

{\bf Literature on reversibility and irreversibility.} 
To be Carnot efficient, systems must be {\it reversible} (create no entropy); for a thermoelectric
there are two requirements for this \cite{Humphrey-Linke2005}. 
Firstly, it must have a  $\delta$-function-like transmission \cite{Mahan-Sofo1996,Kim-Datta-Lundstrom2009,Jordan-Sothmann-Sanchez-Buttiker2013} ($\Delta\big/(\kB T_{L,R}) \to 0$ in Fig~\ref{Fig:T-functions}b) for which the figure of merit $ZT \to \infty$.
Secondly, the load resistance must be such  
that $\eminus V = \eps_0 (1-T_R/T_L)$ \cite{Humphrey-Linke2005}, 
so the reservoirs' occupations are equal at $\eps_0$.  
However, then the power output vanishes, $P_{\rm gen} \propto \Delta^2 \to 0$.

Larger $P_{\rm gen}$ requires heat engines which are {\it irreversible} 
(create a finite amount of entropy per unit of work provided).
The authros of Ref.~\cite{Esposito2009-thermoelec}, motivated by works on classical pumps 
\cite{Curzon-Ahlborn1975,Esposito-PRL2009,Schmiedl-Seifert2008}, proposed increasing 
$P_{\rm gen}$ by keeping $\Delta \to 0$ ($ZT\to \infty$) but choosing the load to
maximize $P_{\rm gen}$, rather than achieve reversibility. 
The resulting Curzon-Alhborn  efficiency
is significantly below $\eta_{\rm eng}^{\rm Carnot}$, yet $P_{\rm gen} \propto \Delta$ remains very small.  
Other works on finite power include   
Refs.~\cite{Nakpathomkun-Xu-Linke2010,Leijnse2010,Meden2013,Hershfield2013}.

Here, we get an efficiency higher than the Curzon-Alhborn efficiency found in 
Ref.~\cite{Esposito2009-thermoelec} for the same (or much larger) $P_{\rm gen}$ 
by making $\Delta$ finite (thereby decreasing $ZT$). 
Thus $ZT \to \infty$ does not give maximal efficiency at given (finite) power output. 
That said, our work does not consider $ZT$ further, as it has little meaning outside the linear response regime
\cite{Zebarjadi2007,Grifoni2011,2012w-pointcont,Meair-Jacquod2013}.

\begin{figure}
\includegraphics[width=\columnwidth]{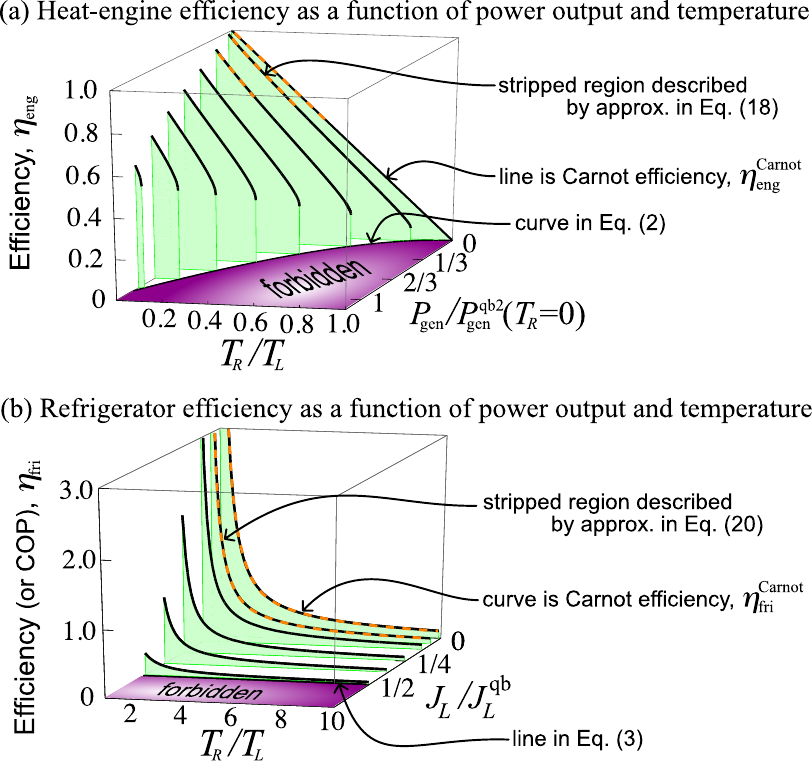}
\caption{\label{Fig:allpowers}
Efficiencies of (a) heat engines and (b) refrigerators. 
The black lines are the maximum allowed efficiency for 
various power outputs as functions of $T_R/T_L$.  
The light green regions indicate allowed efficiencies for those power outputs.
}
\end{figure}

{\bf Heat-engine.}
Here, we find 
the transmission function ${\cal T}^{\mu\mu}_{RL}(\eps)$ 
that maximizes the heat engine efficiency, $\eta_{\rm eng}(P_{\rm gen})=P_{\rm gen}/J_L$,
for a {\it given} power generated, $P_{\rm gen}$.
We treat ${\cal T}^{\mu\mu}_{RL}(\eps)$
as a set of slices as in Fig.~\ref{Fig:T-functions}a,  and find
optimal values of each slice and of the bias, $V$, 
under the constraint of fixed $P_{\rm gen}$. 
A little algebra shows that  $\eta_{\rm eng}(P_{\rm gen})$ will only grow with increasing $\tau^{\mu}_\gamma$, if $\eps_\gamma$ satisfies
\begin{eqnarray}
\left[\eps_\gamma -  \mu \eminus V J'_L/ P'_{\rm gen} \right] \times
\left.{\partial P_{\rm gen}/ \partial \tau^{\mu}_\gamma }\right|_V <0,
\label{Eq:condition}
\end{eqnarray}  
where the prime indicates  $\partial/\partial V$ for fixed ${\cal T}^{\mu\mu}_{RL}(\eps)$.
From this, the optimal ${\cal T}^{\mu\mu}_{RL}(\eps)$ is a boxcar function (Fig.~\ref{Fig:T-functions}b),
\begin{eqnarray}
{\cal T}^{\mu\mu}_{RL}(\eps)
\! &=& \! \left\{ \! \begin{array}{cl} 
N & \hbox{ for }  \mu=1 \ \hbox{ \& } \  \  \eps_0 \! <\! \eps \! 
<\! \eps_1  \phantom{\big|}   
\\
0 & \hbox{ otherwise }  \phantom{\big|} \end{array} \right. \quad 
\label{Eq:boxcar}
\end{eqnarray} 
where $N$ is given in Fig.~\ref{Fig:thermocouple}c.
Then the integrals in Eqs.~(\ref{Eq:JL},\ref{Eq:Pgen}) 
are sums of terms containing logarithmic and dilogarithm functions
of $\eps_0$ and $\eps_1$.
Eq.~(\ref{Eq:condition}) gives
\begin{eqnarray}
\eps_0 = \eminus V  \big/ (1-T_R/T_L), \quad
\eps_1 =  \eminus V \,  J'_L / P'_{\rm gen}.
\label{Eq:eng-bounds}
\end{eqnarray}
Since $J_L$ and $P_{\rm gen}$ depend on $\eps_1$, the second equality is a transcendental equation for $\eps_1$.
Solving this, we get $J_L(V)$ and $P_{\rm gen}(V)$, and so $\eta(V)$.
To get $\eta(P_{\rm gen})$ from $\eta(V)$, we 
invert $P_{\rm gen}(V)$ and substitute for $V$.
Below, we do these steps analytically
for  high power ($P_{\rm gen} =P_{\rm gen}^{\rm QB2}$)
and low power ($P_{\rm gen} \ll P_{\rm gen}^{\rm QB2}$).
For other cases, a numerical solution is plotted in 
Figs.~\ref{Fig:Delta+V}b and \ref{Fig:allpowers}a.

The {\it quantum bound} on 
power output, given in Eq.~(\ref{Eq:P-qb2}), is found by noting that the maximum occurs 
when $P'_{\rm gen}=d P_{\rm gen}/d\eps_1 = 0$.
For this $\eps_1 \to \infty$, so ${\cal T}_{RL}^{\mu\mu}(\eps)$ is a 
Heaviside $\theta$-step function, while $\e^{-\eps_0/(\kB T_L)} \simeq 0.318$.
The efficiency at this maximum power is
\begin{eqnarray}
\eta_{\rm eng} (P_{\rm gen}^{\rm QB2}) &=& \eta_{\rm eng}^{\rm Carnot}\big/ \big( 1+C_0 (1+T_R/T_L) \big),
\end{eqnarray}
with 
$C_0 \simeq 0.936$, so it is always more than $0.3\,\eta_{\rm eng}^{\rm Carnot}$.

For low power output, one can
take Eqs.~(\ref{Eq:JL},\ref{Eq:Pgen}) with Eq.~(\ref{Eq:boxcar}), and easily perform a small $\Delta$ expansion 
up to order $(\Delta/\kB T_R)^3$.
In this limit, Eq.~(\ref{Eq:eng-bounds}) is satisfied by $\eps_0 = 3.2436 \kB T_L$.    
Similarly, taking $P_{\rm gen}$ to lowest order in $\Delta$, 
we rewrite $\eta$ in terms of 
$P_{\rm gen}$ to get Eq.~(\ref{Eq:eta-eng-small-Pgen}).


{\bf Refrigerator.}
A refrigerator's efficiency, or coefficient of performance (COP), 
is $\eta_{\rm fri}(J_L)=J_L/P_{\rm abs}$, where $P_{\rm abs}=-P_{\rm gen}$ is the 
electrical power absorbed by the refrigerator.
We maximize $\eta_{\rm fri}(J_L)$,
for {\it given} cooling power, $J_L$. 
This gives the boxcar function in Eq.~(\ref{Eq:boxcar})
with 
\begin{eqnarray}
\eps_0 = -\eminus V   \,J'_L / P'_{\rm abs},
\quad
\eps_1 =  {-\eminus V  \big/ (T_R/T_L-1)},
\label{Eq:fri-bounds}
\end{eqnarray}
so $\eps_0$ is given by a transcendental equation.
This is solved below analytically at high and low $J_L$,
otherwise the numerical results are given in Figs.~\ref{Fig:Delta+V}c and \ref{Fig:allpowers}b.

The {\it quantum bound} on cooling power in Eq.~(\ref{Eq:J-qb-fri}),
occurs when the transmission function is a 
$\theta$-step function ($\eps_0=0$, $\eps_1 \to \infty$ and $-\eminus V \to \infty$).
Then $\eta_{\rm fri}(J_L)$ is zero, since $V$ is infinite.
However one gets exponentially close to this limit for
$-\eminus V \gg \kB T_R$, for which $\eta_{\rm fri}(J_L)$ is finite
(see  Fig.~\ref{Fig:allpowers}b).
In the opposite limit, $J_L \ll J_L^{\rm QB}$, 
an expansion up to third order in $\Delta/(\kB T_L)$ gives Eq.~(\ref{Eq:eta-fri-smallJ}).


{\bf Phonons and photons.}
These unavoidably carry heat from hot to cold in parallel with the electronic flow. 
Their heat current, $J_{\rm ph}$, depends nonlinearly on $T_{L,R}$ (given by a Stefan-Boltzmann law 
or similar \cite{Pendry1983,photons,phonons}).
Then maximal efficiencies for heat engines and refrigerators are suppressed, and given by 
\begin{eqnarray}
\eta_{\rm eng}^{\rm e+ph} (P_{\rm gen}) 
&=&
\big[ \eta_{\rm eng}^{-1} (P_{\rm gen}) +J_{\rm ph} / P_{\rm gen} \big]^{-1},  
\nonumber
\\
\eta_{\rm fri}^{\rm e+ph}(J-J_{\rm ph}) 
&=&
\big(1-J_{\rm ph}/J\big)\, \eta_{\rm fri}(J)  \ \hbox { for } J> J_{\rm ph},
\nonumber
\end{eqnarray}
where $\eta_{\rm eng} (P_{\rm gen})$, $\eta_{\rm fri} (J)$ are efficiencies at $J_{\rm ph}=0$.

In many devices, there is a large phonon or photon heat flow, $J_{\rm ph}$.
For a heat engine in a situation where  $J_{\rm ph} \gg \eta_{\rm eng} P_{\rm gen}$, one has 
$\eta_{\rm eng}^{\rm e+ph} (P_{\rm gen}) 
= P_{\rm gen}/J_{\rm ph}$.  Thus, the efficiency is maximal when the power is maximal, as given by Eq.~(\ref{Eq:P-qb2}).
For a refrigerator to cool, it needs $J-J_{\rm ph}>0$, so 
one may need the maximum cooling power in Eq.~(\ref{Eq:J-qb-fri}) when $J_{\rm ph}$ is large.
In both cases, this corresponds to a $\theta$-step transmission function.


{\bf Inelastic effects.}
Inelastic electron-phonon and electron-electron interactions in the quantum system 
are not accounted for in the above theory.
However they will be negligible if the quantum system is small enough.
At 700 Kelvin, electrons typically travel tens of nanometres before an inelastic scattering,
so if the quantum system is a few Angstroms across, inelastic effects may be insignificant.  Below 1 Kelvin, electrons can traverse micro-sized structures without inelastic scattering.
We will address inelastic effects in detail elsewhere \cite{w2014} 
using a voltage-probe model \cite{voltage-probe}.  
We will show that they {\it cannot} increase the maximum power beyond 
Eqs.~(\ref{Eq:P-qb2},\ref{Eq:J-qb-fri}).  
For low powers, we will show that they {\it cannot} increase the maximum efficiencies beyond
Eqs.~(\ref{Eq:eta-eng-small-Pgen},\ref{Eq:eta-fri-smallJ}).
For intermediate powers, it remains 
open whether they could raise the maximum efficiency, 
however there is no reason to think so.

 
{\bf Many quantum systems in parallel.}
Increasing the number of modes, $N$, increases the efficiency at given power output.
This is because the quantum bounds in Eqs.~(\ref{Eq:P-qb2},\ref{Eq:J-qb-fri}) go like $N$,
and the efficiency goes toward Carnot efficiency as these bounds grow.   

However most thermoelectric quantum systems 
have $N \sim1$ (exceptions being SNS structures \cite{Pekola-reviews}).
Then large $N$ would require many $N=1$ systems in parallel.
For a surface covered with a certain density of such systems 
\cite{Jordan-Sothmann-Sanchez-Buttiker2013}, 
Eqs.~(\ref{Eq:P-qb2},\ref{Eq:J-qb-fri}) become
bounds on the power per unit area.  
Carnot efficiency is only approachable when the power per unit area 
is much less than these bounds.
The number of modes per unit area cannot exceed $\lambda_{\rm F}^{-2}$, for Fermi wavelength 
$\lambda_{\rm F}$.  Thus, Eq.~(\ref{Eq:P-qb2}) tells us that to get 100 W of power output from a semiconductor thermoelectric {\color{red}(with $\lambda_{\rm F}\sim 10^{-8}$m) }
between reservoirs at 700 K and 300 K, one needs a cross section of at least  {\color{red}4 mm$^2$}.  
To get this power at 90\% of Carnot efficiency, one needs a cross section of  at least  {\color{red}0.4 cm$^2$}.
Remarkably, it is {\it quantum mechanics} which gives these bounds, 
even though the cross sections in question are macroscopic.

\begin{figure}[t]
\includegraphics[width=0.9\columnwidth]{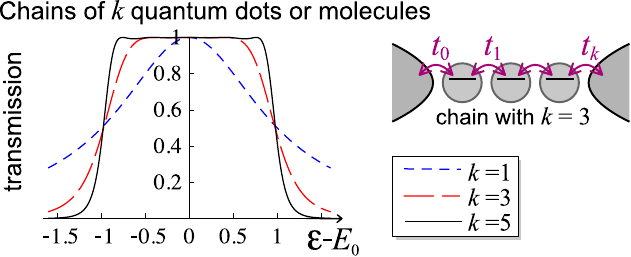}
\caption{\label{Fig:band}
Chain of systems with levels at energy $E_0$ and hoppings $\{t_i\}$, 
will hybridize to form a band centred at $E_0$, with a width given by the hopping.
If the hoppings are smallest at the chain's middle and larger at its ends \cite{w2014},
the transmission becomes increasingly like a boxcar function as one increases $k$
(the bandwidths in the plots have been set to one). }
\end{figure}


{\bf Concluding remark on implementation.}
Whenever $(1-T_R/T_L)$ is not small, the transmission function for a system must be found 
self-consistently to capture charging effects.  
This work has shown that maximum efficiency for given power output occurs when this transmission is a boxcar function (with correct position and width).
Fig.~\ref{Fig:band} shows one potential implementation, the energy levels should be chosen so that they align at energy $E_0$ when the optimal bias is applied.
To get maximum power output is simpler, since a  good point contact has the required 
$\theta$-step function transmission.

{\bf Acknowledgements.} I thank M.~B\"uttiker for proposing a band as a candidate for the
boxcar transmission.




\begin{thebibliography}{1}


\bibitem{QuantumThermodyn-book}
J.~Gemmer, M.~Michel, and G~ Mahler,
{\it Quantum Thermodynamics}
(Springer-Verlag, Berlin, 2010).


\bibitem{books}
H.J.~Goldsmid, {\it Introduction to Thermoelectricity} (Springer, Heidelberg, 2009). 
\bibitem{DiSalvo-review}
F.J.~DiSalvo, Science {\bf 285}, 703 (1999).
\bibitem{Shakouri-reviews}
A.~Shakouri and M.~Zebarjadi, 
Chapt 9 of {\it Thermal nanosystems and nanomaterials}, S.~Volz (Ed.)  (Springer, Heidelberg, 2009).
A.~Shakouri, Annu. Rev. Mater. Res. {\bf 41}, 399 (2011).

\bibitem{Pekola-reviews}
F.~Giazotto, T.T.~Heikkila, A.~Luukanen, A.M.~Savin, J.P.~Pekola,
Rev.~Mod.~Phys.~{\bf 78}, 217 (2006).
J.T.~Muhonen, M.~Meschke, J.P.~Pekola,  Rep.~Prog.~Phys.~{\bf 75}, 046501 (2012).

\bibitem{Paulson-Datta2003}
M.\ Paulsson and S.\ Datta,
Phys. Rev. B {\bf 67}, 241403(R) (2003).

\bibitem{Reddy2007}
P.\ Reddy, S.Y.\ Jang, R.A.\ Segalman, and A.\ Majumdar,
Science, {\bf 315}, 1568 (2007).

\bibitem{Pendry1983}
J.B.~Pendry, J.~Phys.~A.: Math.\ Gen.\ {\bf 16}, 2161 (1983).



\bibitem{Mahan-Sofo1996}
G.D.~Mahan and J.O.~Sofo, Proc.\ Nat.\ Acad.\ Sci.\ U.S.A. {\bf 93}, 7436 (1996).

\bibitem{Humphrey-Linke2005}
T.E.~Humphrey and H.~Linke, 
Phys.\ Rev.\ Lett.\ {\bf 94}, 096601 (2005).

\bibitem{Kim-Datta-Lundstrom2009}
R.~Kim, S.~Datta, M.S.~Lundstrom,
J.~Appl.~Phys.\ {\bf 105}, 034506 (2009).

\bibitem{Jordan-Sothmann-Sanchez-Buttiker2013}
A.N.\ Jordan, B.\ Sothmann, R.\ Sanchez, and M.\ B\"uttiker,
Phys.~Rev.~B, {\bf 87}, 075312 (2013). 

\bibitem{Reichl-book}
see e.g.~chapter 3 of L.E.~Reichl, {\it A Modern Course in Statistical Physics} 3rd edition 
(Wiley-VCH, Weinheim, 2009).

\bibitem{footnote:2012w-2ndlaw}
This is $6A_0(1-T_R/T_L)$ smaller than Ref.~\cite{2012w-2ndlaw}'s bound, 
which came from taking $J_{\rm L}=J_{L}^{\rm QB}$ and $\eta_{\rm eng}= \eta_{\rm eng}^{\rm Carnot}$.  
However, no two-lead system can satisfy both at once. 

\bibitem{Footnote:bulk-semicond}
Linear response fails when the ratio of temperature drop, $\delta T$, to temperature, $T$, is significant 
on the scale of the electron relaxation length $l_{\rm rel}$.  
In bulk systems ($L\gtrsim 1$mm, $l_{\rm rel}\lesssim$ 0.1$\mu$m), 
linear response \cite{Mahan-Sofo1996} works well, since the
ratio is $\Delta T/T \times l_{\rm rel}/L$,
which is small even for $\Delta T/T \sim 1$.
For quantum systems ($L \ll l_{\rm rel}$), it fails whenever $\Delta T/T$ is not small.


\bibitem{Christen-ButtikerEPL96}
T.~Christen and M.~B\"uttiker, Europhys.~Lett.~{\bf 35}, 523 (1996). 

\bibitem{Sanchez-Buttiker}
D.~Sanchez and M.~B\"uttiker, Phys.~Rev.~Lett.~{\bf 93}, 106802 (2004).

\bibitem{Sanchez-Lopez2013}
D.~Sanchez and R.~Lopez, Phys. Rev. Lett. {\bf 110}, 026804 (2013).
R. L\'opez, and D.~S\'anchez, Phys.\ Rev.\ B {\bf 88}, 045129 (2013).

\bibitem{2012w-pointcont}
R.S.~Whitney,  Phys.\ Rev.\ B {\bf 88}, 064302 (2013).

\bibitem{Meair-Jacquod2013}
J.~Meair, and Ph.~Jacquod, J. Phys.: Condens. Matter {\bf 25} 082201, (2013). 
 

\bibitem{Galperin2007-2008}
M.~Galperin, M.A.~Ratner, and A.~Nitzan,
J.\ Phys.: Condens. Matter {\bf 19}, 103201 (2007).
M.~Galperin,  A.~Nitzan, and M.A.~Ratner, Mol.\ Phys.\ {\bf 106}, 397 (2008).

\bibitem{Murphy2008}
P.~Murphy, S.~Mukerjee, and J.~Moore, 
Phys.\ Rev.\ B {\bf 78}, 161406(R) (2008).


\bibitem{Nakpathomkun-Xu-Linke2010}
N.~Nakpathomkun, H.Q.~Xu, and H.~Linke,
Phys.\ Rev.\ B, {\bf 82}, 235428 (2010).


\bibitem{Esposito2009-thermoelec}
M.\ Esposito, K.\ Lindenberg and C.\ Van den Broeck, Europhys.~Lett. {\bf 85},  60010 (2009).


\bibitem{Meden2013}
D.M.\ Kennes, D.\ Schuricht and V.\ Meden, 
Europhys.~Lett. {\bf 102} (2013) 57003 (2013).



\bibitem{2012w-2ndlaw}
R.S.~Whitney,  Phys.~Rev.~B {\bf 87}, 115404 (2013). 



\bibitem{Bruneau2012}
L. Bruneau, V. Jak\v{s}i\'{c}, and C.-A. Pillet, Commun. Math. Phys. {\bf 319}, 501 (2013).




\bibitem{Butcher1990}
P.N.~Butcher, J.~Phys.: Condens.~Matter,  {\bf 2}, 4869 (1990). 

\bibitem{Molenkamp1992} 
L.W.~Molenkamp, Th.~Gravier, H.~ van Houten, O.J.A.~Buijk, M.A.A.~Mabesoone, and C.T.~Foxon,
Phys.~Rev.~Lett.~{\bf 68}, 3765 (1992).

\bibitem{Vavilov-Stone2005} 
M.G.\ Vavilov and A.D.\ Stone, Phys.~Rev.~B {\bf 72}, 205107 (2005).



\bibitem{Zebarjadi2007}
M.~Zebarjadi, K.~Esfarjani, and A.~Shakouri, Appl. Phys. Lett. {\bf 91}, 122104 (2007);
MRS Proceedings 1044 U10-04 (2008).

\bibitem{Nozaki2010}
D.~Nozaki, H.~Sevin\c cli, W.~Li, R.~Guti\'errez, and G.~Cuniberti,  
Phys.~Rev.~B {\bf 81}, 235406 (2010). 


\bibitem{jw-epl}
Ph.~Jacquod and R.S.~Whitney, Europhys.~Lett.~{\bf 91}, 67009 (2010).

\bibitem{Saha2011}
K.K.~Saha, T.~Markussen, K.S.~Thygesen, and B.K.~Nikoli\'c, Phys.~Rev.~B {\bf 84}, 041412(R) (2011).

\bibitem{Karlstrom2011}
O. Karlstr\"om, H. Linke, G. Karlstr\"om, and A. Wacker,  Phys.~Rev.~B {\bf 84}, 113415 (2011). 


\bibitem{jwmb}
Ph.~Jacquod, R.S.~Whitney, Jonathan Meair, and M.~B\"uttiker, 
Phys. Rev. B {\bf 86}, 155118 (2012).


\bibitem{Hershfield2013}
S.\ Hershfield, K.A.\ Muttalib, and B.J.\ Nartowt,  
Phys.~Rev.~B, {\bf 88}, 085426 (2013).



\bibitem{Entin-Wohlman2011}
O.~Entin-Wohlman, Y.~Imry, and A.~Aharony,
Phys.\ Rev.~B, {\bf 82}, 115314 (2010).
\bibitem{SB2011}
R.~S\'anchez and M.\ B\"uttiker Phys.\ Rev.\ B {\bf 83}, 085428 (2011).
\bibitem{SSJB2012}
B.~Sothmann, R.~S\'anchez, A.N.~Jordan, and M.~B\"uttiker,
Phys.~Rev.~B 85, 205301 (2012).
\bibitem{Trocha2012}
P.~Trocha and J.~Barna\'s,  Phys.~Rev.~B {\bf 85}, 085408 (2012).
\bibitem{Brandner2013}
K.\ Brandner, K.\ Saito and U.\ Seifert, Phys.\ Rev.\ Lett.\ {\bf 110}, 070603 (2013).


\bibitem{Curzon-Ahlborn1975}
F.L.~Curzon, and B.~Ahlborn, Am.~J.~Phys.\  {\bf 43}, 22 (1975).

\bibitem{Schmiedl-Seifert2008}
T.~Schmiedl and U.~Seifert, EPL, {\bf 81}, 20003 (2008).

\bibitem{Esposito-PRL2009}
M.~Esposito, K.~Lindenberg, and C.~Van den Broeck, Phys.\ Rev.\ Lett.\
{\bf 102}, 130602 (2009).


\bibitem{Leijnse2010}
M.\ Leijnse, M.R.\ Wegewijs, and K.\ Flensberg,
Phys.\ Rev.\ B {\bf 82}, 045412 (2010).


\bibitem{Grifoni2011}
B.~Muralidharan, M.~Grifoni
Phys.~Rev.~B {\bf 85}, 155423 (2012).

\bibitem{photons}
D.R.\ Schmidt, R.J.\ Schoelkopf, and A.N.\ Cleland, 
Phys.\ Rev.\ Lett.\ {\bf 93}, 045901 (2004). 
L.M.A.~Pascal, H.~ Courtois, and F.W.J.~Hekking,
Phys.\ Rev.~B {\bf 83}, 125113 (2011).

\bibitem{phonons}
J.S.~Heron, T.~Fournier, N.~Mingo, O.~Bourgeois,	
Nano Lett.\ {\bf 9},  1861 (2009).
J.-S.\ Heron, C.\ Bera, T.\ Fournier, N.\ Mingo, and O.\ Bourgeois,
Phys.~Rev.~B 82, 155458 (2010).


\bibitem{w2014}
R.S.~Whitney, in preparation.

\bibitem{voltage-probe}
M.~B\"uttiker, IBM J.\ Res.\ Dev.\ {\bf 32}, 63 (1988).




\end{thebibliography}
\end{document}